\documentclass[sort&compress,final]{aipproc}\layoutstyle{8x11double}
\begin{document}\title{Quark--Hadron Duality and Effective
Continuum Thresholds in Dispersive Sum Rules}
\classification{12.38.-t, 12.38.Lg, 11.10.St, 11.55.Hx, 03.65.Ge}
\keywords{nonperturbative quantum chromodynamics, dispersive sum
rules, quark--hadron duality, effective continuum threshold,
correlator}
\author{Wolfgang LUCHA}{address={Institute for High Energy Physics,
Austrian Academy of Sciences, Nikolsdorfergasse 18, A-1050 Vienna,
Austria}}\author{Dmitri MELIKHOV}{address={Institute for High
Energy Physics, Austrian Academy of Sciences, Nikolsdorfergasse
18, A-1050 Vienna, Austria},altaddress={Faculty of Physics,
University of Vienna, Boltzmanngasse 5, A-1090 Vienna, Austria}}
\author{Silvano SIMULA}{address={INFN, Sezione di Roma
III, Via della Vasca Navale 84, I-00146 Roma, Italy}}\maketitle

Recently, we developed a new procedure for extracting hadron
characteristics from QCD sum rules by proposing to improve the
implementation of `quark--hadron duality' by allowing our
continuum threshold $z$ to depend on both the involved momenta $q$
and a parameter $T$ entering upon application of a {\em Borel
transformation\/}: $z=z(T,q)$ \cite{LMS:PRD76,LMS:YaF2008,
LMS:PLB657,LMS:PLB671,DM:PLB671,LMS:PRD79,LMS:JPG37,LMS:PRD80,
LMS:PLB687,LMS:YaF2010}. The actual behaviour of $z(T,q)$ may be
found from a fit to the (experimentally fixed) hadron masses. Such
modified method enables us to quantify the uncertainty induced by
the {\em assumed\/} quark--hadron duality {\em approximation\/},
and to achieve a significantly higher accuracy of the predictions.

The advantages of such concept are most convincingly demonstrated
for the case of simple quantum-mechanical (QM) models with, e.g.,
a harmonic-oscillator interaction of angular frequency $\omega,$
where luckily for all bound-state features one is interested in
the results are known exactly. This serendipity renders possible
to perform {\em quantitative\/} comparisons of our (unambiguous)
QM answers with the outcomes of the approximate extraction
procedure in use.

\begin{figure}[h]\begin{tabular}{c}
\includegraphics[scale=.512672]{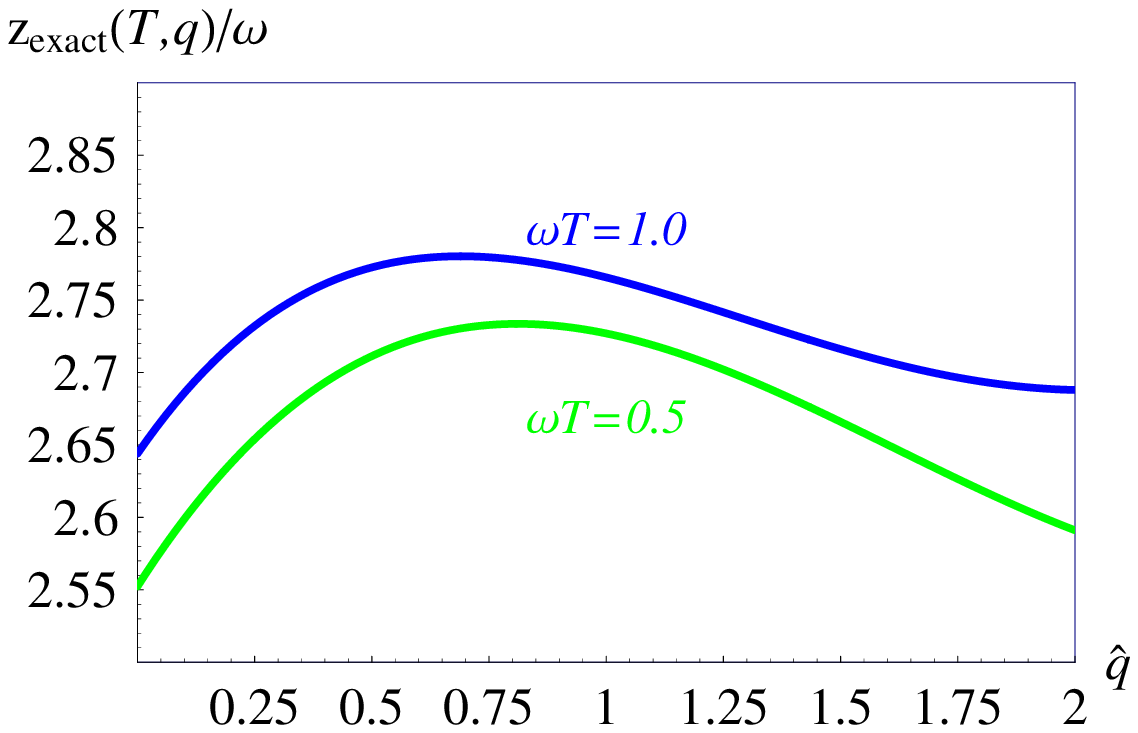}\\
\includegraphics[scale=.512672]{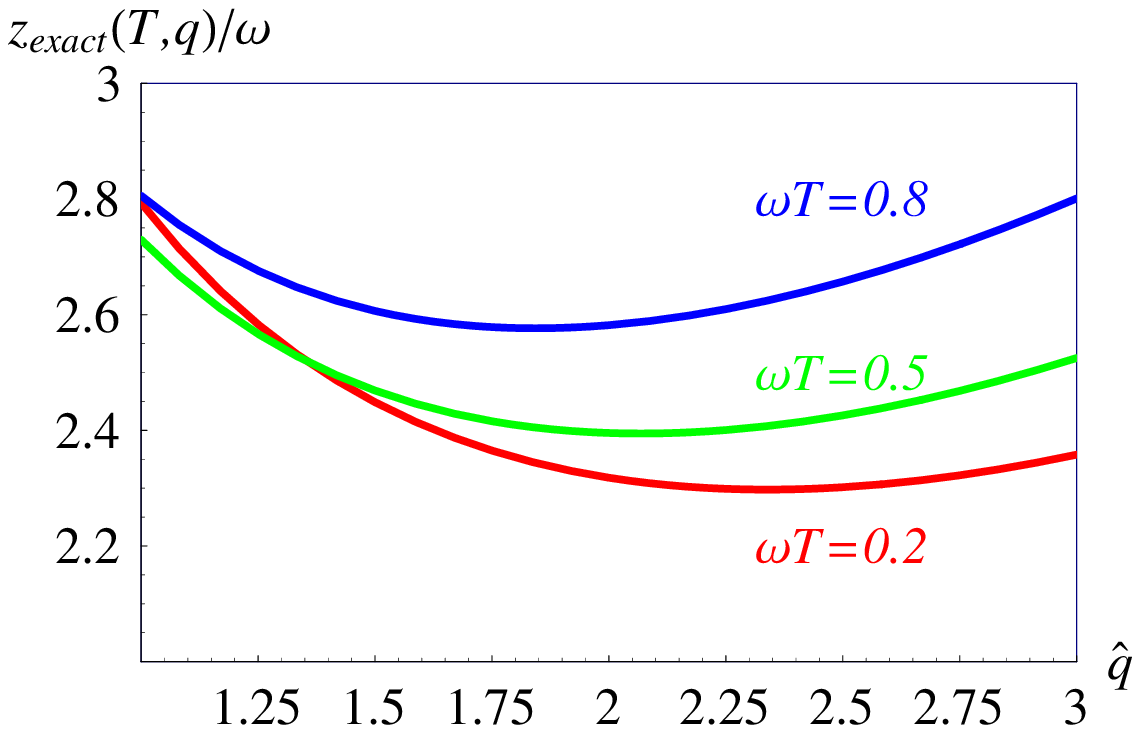}\end{tabular}
\end{figure}

Our main findings are the proof that the exact effective continuum
threshold $z_{\rm exact}$ {\em does\/} depend on both the Borel
parameter and the relevant momenta, and is {\em not\/} universal
but varies with the studied correlator. For the dependence on the
(dimensionless) momentum $\hat q$ the figure illustrates this
issue for the three-current vacuum--vacuum (top) and the
two-current vacuum--hadron (bottom) correlator \cite{LMS:YaF2010}.
Realizing the {\em procedure\/} to be similar in QM and in QCD, as
first {\em QCD application\/} of this idea \cite{LMS:fP} we
successfully predicted the heavy pseudoscalar meson decay
constants.

\begin{theacknowledgments}D.I.M.\ would like to express his
gratitude to the Austrian Science Fund FWF for support under
Project no.~P20573.\end{theacknowledgments}

\bibliographystyle{aipproc}\end{document}